\documentclass[eqsecnum,aps,prd,nofootinbib,preprint]{revtex4}

\usepackage{amsmath}
\usepackage{amssymb}
\usepackage{mathrsfs}
\usepackage{bbm}
\usepackage{graphicx}
\usepackage{color}
\usepackage{pstricks,pst-coil,pst-plot}
\usepackage[normalem]{ulem}

\newrgbcolor{darkred}{0.7 0.13 0.13}

\begin{document}

\title{Flowing Funnels:  Heat sources for field theories and the AdS${}_3$ dual of CFT${}_2$ Hawking radiation}

\author{Sebastian Fischetti\footnote{\tt sfichet@physics.ucsb.edu}
 and Donald Marolf\footnote{\tt marolf@physics.ucsb.edu}\\
 Physics Department, UCSB, Santa Barbara, CA 93106\\
 E-mail: sfischet@physics.ucsb.edu, marolf@physics.ucsb.edu}

\author{Donald Marolf}

\affiliation{Department of Physics \\ University of California,
Santa Barbara \\ Santa Barbara, CA 93106, USA }

\begin{abstract}
We construct the general 2+1 dimensional asymptotically AdS${}_3$ solution dual to a stationary 1+1 CFT state on a black hole background.  These states involve heat transport by the CFT between the 1+1 black hole and infinity (or between two 1+1 black holes), and so describe the AdS dual of CFT Hawking radiation.
Although the CFT stress tensor is typically singular at the past horizon of the 1+1 black hole, the bulk 2+1-dimensional solutions are everywhere smooth, and in fact are diffeomorphic to AdS${}_3$.  In particular, we find that Unruh states of the CFT on any finite-temperature 1+1 black hole background are described by extreme horizons in the bulk.
\end{abstract}

\date{\today}

 \maketitle

\tableofcontents

\section{Introduction}
\label{intro}

The study of field theories far from equilibrium is a classic problem of long-standing interest.  While much can be learned from perturbation theory, more complete results are difficult to obtain.   As is by now well known, gauge/gravity duality can be a useful tool to study non-perturbative effects.  There has thus been significant interest in using this framework to study plasmas with strong time dependence and the approach to thermalization (see e.g. \cite{Chesler:2010bi,Heller:2011ju,Bantilan:2012vu} for recent examples and further references), though it has mostly been used to study the equilibrium properties of field theories, or perhaps small perturbations away from equilibrium.  Below, we use this duality to study heat transport far from equilibrium in a strongly coupled large N CFT.

We are interested in particular in the response of the field theory when coupled to heat sources or sinks at finite locations.  For the purposes of this section, we consider a field theory in $d$ spacetime dimensions. A convenient way to introduce such sources is to place the CFT on a background non-dynamical spacetime containing black holes with surface gravity $\kappa$, which have temperatures $T_{BH} = \kappa/2\pi$ due to the Hawking effect.  The problem of heat transport then becomes one of computing the expectation value of the stress tensor in the given background.  Note that, since gravity is not dynamical in this context, we can choose the metric at will.  In particular, we can include as many black holes as we like at locations of our choosing, and we are free to assign their surface gravities as desired.  One then seeks states of the field theory which are smooth across all future horizons.  Stationary such states are analogues of the Unruh vacuum \cite{Unruh:1976db} for black holes in asymptotically flat spacetimes.

Gauge/gravity duality for large $N$ field theories \cite{Maldacena:1997re} has been used to study related settings
in \cite{AM,Tanaka:2002rb,Emparan:2002px,Wiseman:2001xt,Wiseman:2002zc,Casadio:2002uv,Karasik:2003tx,Kudoh:2003xz,Kudoh:2003vg,Kudoh:2004kf,Karasik:2004wk,Fitzpatrick:2006cd,Yoshino:2008rx,Gregory:2008br,
Hubeny:2009ru,Hubeny:2009kz,Hubeny:2009rc,Kleihaus:2011yq,Figueras:2011va,Figueras:2011gd,Caldarelli:2011wa}.  Though this exploration involved certain tensions and subtleties, the picture that emerged in \cite{Hubeny:2009ru} (building on \cite{Fitzpatrick:2006cd}) is one with two important phases for each black hole.  In the so-called ``funnel phase'' a given black hole exchanges heat with distant regions much as in a free theory with a similar number of fields.  One may say that grey body factors are ${\cal O}(1)$ even at large $N$.  But in the contrasting ``droplet phase'' there is    no conduction of heat between a given black hole and the region far away at leading order in large $N$.   In effect, all grey body factors associated with the black hole vanish at this order.\footnote{To be more precise, the grey body factors are non-zero only for a number of degrees of freedom that scales like $N^0=1$.}   While we save a more complete review for section \ref{review}, we mention that the terms ``droplet'' and ``funnel'' refer to the shape of the bulk horizon in the dual gravitational theory; see especially figure \ref{f:fundrop}.  Additional phases are also possible that conduct heat between subsets of nearby black holes but not to infinity; these are of less concern below.  We also mention that 1+1 CFTs (and their 2+1 AdS duals) are a special case in which only the funnel phase is allowed.

The above works have focussed on cases without heat flow; i.e., either droplets (in which heat does not flow at leading order) or on equilibrium funnels.  But heat transport is an interesting phenomenon and, moreover, at least for $d \ge 3$ the corresponding solutions of the dual AdS${}_{d+1}$ gravitational theory are expected to have novel properties.  For example, time-independent such solutions should be black holes which (in some sense) have a temperature that varies along the horizon.  But there is no generally accepted definition of horizon temperature which allows the temperature to vary\footnote{Except of course within the domain of the gradient expansion, as in the fluid-gravity correspondence \cite{Bhattacharyya:2008jc};  see also  \cite{Kinoshita:2011qs}.  For proposals in more general contexts see e.g. \cite{Abreu:2010ru} for a recent paper and references.  If flowing funnels could be constructed, it might be interesting to apply these proposals and study the results.}.   Indeed, the fact that any definition of temperature should vary implies that the horizon is not a Killing horizon, which is already a novel property for a stationary solution. There is also an interesting question of whether such black holes should have a regular past horizon.  On the one hand, from the field theory point of view, the CFT stress tensor must diverge on the past horizon of the (now fixed and non-dynamical) black hole metric in any state with both heat flow and time-translation symmetry.  Ref. \cite{Hubeny:2009ru} thus expected the past horizon of the dual bulk solutions to be singular in the presence of (stationary) heat flow.  But cases are known where similarly singular stress tensors are described by smooth bulk solutions \cite{Headrick:2010zt,Marolf:2010tg,Hung:2011nu}. Perhaps this could be the case here as well.

This paper provides a first step in this direction by constructing examples of such `flowing funnel' solutions in AdS${}_3$, which are then dual to heat transport in a $1+1$ CFT; i.e., for the case $d=2$.   The behavior of the stress tensor (and thus of heat transport) in $1+1$ CFT's is fully determined by conformal invariance, and of course the same is true of the AdS${}_3$ description.  This is therefore not an example where one expects gauge/gravity duality to lead to significant new insights for the field theory.  In addition, a consequence of this conformal symmetry is that states are characterized by {\it two} (left- and right-moving) temperatures $T_L,T_R,$ both of which are necessarily constant (independent of spatial position) in a stationary solution.  So this is also not a context where we will learn about bulk horizons with non-constant temperature.  Nevertheless, it gives a prime opportunity to see how heat transport in the CFT is encoded in the dual bulk solutions.  We may also hope that the explicit solutions given below will provide a useful starting point for studying the higher-dimensional case.

We begin with a brief review of black funnels, black droplets, and their dual field theory description in section \ref{review}.
Section \ref{BTZ} then constructs flowing funnels from rotating BTZ black holes. We close with some further discussion in section \ref{disc} and relegate details of the Fefferman-Graham coordinates to an appendix.

\section{A brief review of droplets and funnels}
\label{review}

Studies of the above framework using gauge/gravity duality have led to both surprising phenomena and significant controversy.  The first such study appears to have been \cite{AM}, which used the AdS${}_4$ C-metric \cite{Plebanski:1976gy} to find a bulk solution describing the dual CFT on an asymptotically flat black hole spacetime.  The surprise was that, at leading order in large N, the CFT stress tensor was completely static and described no flow of heat from the (finite temperature) black hole to infinity (where the state approached the vacuum).  Indeed, the expectation of finite heat flow was so strong that it motivated predictions \cite{Tanaka:2002rb,Emparan:2002px} (see also \cite{Yoshino:2008rx,Kleihaus:2011yq})) that black holes on Randall-Sundrum brane-worlds \cite{Randall:1999ee,Randall:1999vf} could not be stationary even at the classical level\footnote{The brane in a Randall-Sundrum brane-world spacetime can be thought of as a boundary for an asymptotically AdS spacetime which has been placed at finite distance and given a boundary condition that makes the boundary metric dynamical.  At least roughly speaking, this makes the system dual to a field theory coupled to dynamical gravity.  Hawking radiation in the dual field theory would therefore cause the brane black hole to shrink.  If this effect occurs at leading order in large N, then it would be visible at the classical level from the bulk point of view.}, in contradiction with the natural intuition based on gravity in the bulk (see \cite{Chamblin:1999by} for details).  The difficulty in finding such stationary solutions with black holes (see e.g. \cite{Wiseman:2001xt,Wiseman:2002zc,Kudoh:2003xz,Kudoh:2003vg,Kudoh:2004kf,Chamblin:2000ra,Casadio:2002uv,Karasik:2003tx,Karasik:2004wk}) made these arguments seem compelling for some time, though modern numerical techniques have established that these solutions do in fact exist \cite{Figueras:2011va,Figueras:2011gd}.  As in \cite{AM}, although the field theory contributes a non-trivial stress tensor one finds no transport of energy to infinity at leading order in $N$; i.e., at ${\cal O}(N^2)$ for \cite{Figueras:2011va,Figueras:2011gd}.\footnote{Here and below we will count powers of $N$ as appropriate to a large $N$ SU($N$) Yang-Mills theory.  This in particular describes theories dual to bulk spacetimes asymptotic to AdS${}_5 \times X$ for compact 5-dimensional manifolds $X$.}  Indeed, from the bulk viewpoint, heat transport can occur only due to a quantum process (bulk Hawking radiation) which is an effect of order $N^0=1$.  See also \cite{Gregory:2008br}.

While intuition from weak coupling may make this tiny heat transport seem surprising,  Fitzpatrick, Randall and Wiseman \cite{Fitzpatrick:2006cd} pointed out that similar phenomena are in fact well-known at strong coupling and large $N$.  In particular, they noted that confined phases of large $N$ gauge theories have conductivities of order $1$ and not of order $N^2$.  Though the above theories are not strictly confining, we see that an effect of this magnitude would explain the results of \cite{Figueras:2011va,Figueras:2011gd}.

A somewhat more complete picture was constructed in \cite{Hubeny:2009ru} and explored further in \cite{Hubeny:2009kz,Hubeny:2009rc,Caldarelli:2011wa}.  The basic approach of \cite{Hubeny:2009ru} was to i) explicitly add a thermal bath at infinity, ii) follow the natural intuition for the bulk gravitating solutions, and iii) translate the results into an explanation\footnote{Here we mean a self-consistent scenario for the behavior of the strongly-coupled large N gauge theory.  A full explanation would of course require the scenario to be derived from first principles, but this remains an open problem.} of the effect in the gauge theory.
The rest of this section briefly summarizes the arguments of \cite{Hubeny:2009ru} with minor additions and clarifications.

Let us take the gauge theory to be conformal, and to be deconfined at all temperatures $T > 0$. We take the theory to live on some asymptotically flat spacetime and imagine coupling the system to a large heat bath (at some temperature $T_\infty$) far from the black hole. It is clear that this heat bath should fill the spacetime with a thermal plasma -- at least far from the black hole where the spacetime is nearly flat.  In order to discuss the dual gravitational solution, it is useful to introduce two coordinates: $r$, which parametrizes the distance from the black hole in the gauge theory and $z$, chosen so that the gauge theory `lives' on the AdS boundary at $z=0$ and which parametrizes the distance into the bulk.  To be concrete, at least near the boundary one might take these to be part of a Fefferman-Graham type coordinate system\footnote{Though there is no a priori guarrantee that such coordinates are regular across the full bulk horizon. See e.g. the discussion of 2+1 funnels in \cite{Hubeny:2009ru}.} in which the bulk metric takes the form

\begin{equation}
ds_{d+1}^2 = z^{-2} \left( dz^2 + g_{ij} dx^i dx^j \right),
\end{equation}
with $r$ being one of the boundary coordinates $x^i$ and $ds^2_d = g_{ij} dx^i dx^j$ being the asymptotically-flat black hole spacetime (which we call the `boundary black hole' or equivalently the `gauge theory black hole') on which the gauge theory is to be studied.  One expects the thermal plasma to be described by a bulk horizon that approximates that of the familiar planar Schwarzschild black hole at large $r$.  On the other hand, one also expects the horizon of the boundary black hole to extend into the bulk.  There are then two natural classes of possible bulk solutions.  If the above two horizons connect to form a single smooth horizon, the solution is said to describe a ``black funnel."    If they are instead disconnected, the solution describes a ``black droplet, suspended above a (deformed) planar black hole."  These two situations are sketched in figure 1.

\begin{figure}[t]
\begin{center}
\hfill
\begin{pspicture}(-1,0)(16,5)
\psset{unit=0.95cm}


\pscustom{
\gsave
	\pscurve(0,2)(0.8,2.05)(1.2,2.2)(1.6,2.6)(2,4)
	\psline(5,4)
	\pscurve(5.4,2.6)(5.8,2.2)(6.2,2.05)(7,2)
	\psline(7,1.2)(0,1.2)(0,2)
	\fill[fillstyle=solid,fillcolor=lightgray]
\grestore
}

\psline(0,4)(7,4)
\psdots(2,4)(5,4)
\pscurve(0,2)(0.8,2.05)(1.2,2.2)(1.6,2.6)(2,4)
\pscurve(5,4)(5.4,2.6)(5.8,2.2)(6.2,2.05)(7,2)

\psline{|-|}(2,4.2)(5,4.2)
\rput[b](3.5,4.3){$2R$}
\psline{|-|}(-0.2,2)(-0.2,4)
\rput[r](-0.3,3){$T_\infty^{-1}$}

\psline{->}(6,3.4)(5.2,3.7)
\rput[l](6.2,3.4){Throat}
\psline{->}(2.6,1.6)(1.4,2.1)
\psline{->}(4.4,1.6)(5.6,2.1)
\rput[t](3.5,1.6){Shoulders}
\rput[b](0.9,4.05){Boundary}
\rput(3.5,3){Black Funnel}
\rput(3.5,0.5){$(a)$}


\pscustom{
\gsave
	\pscurve(12,4)(12.1,3.8)(12.3,3.6)(12.5,3.55)(12.7,3.6)(12.9,3.8)(13,4)
	\psline(12,4)
	\fill[fillstyle=solid,fillcolor=lightgray]
\grestore
}

\pscustom{
\gsave
	\pscurve(9,2)(10,2)(11,2)(12.5,2.2)(14,2)(15,2)(16,2)
	\psline(16,1.2)(9,1.2)
	\fill[fillstyle=solid,fillcolor=lightgray]
\grestore
}

\psline(9,4)(16,4)
\psdots(12,4)(13,4)
\pscurve(12,4)(12.1,3.8)(12.3,3.6)(12.5,3.55)(12.7,3.6)(12.9,3.8)(13,4)
\pscurve(9,2)(10,2)(11,2)(12.5,2.2)(14,2)(15,2)(16,2)

\psline{|-|}(12,4.2)(13,4.2)
\rput[b](12.5,4.3){$2R$}
\psline{|-|}(8.8,2)(8.8,4)
\rput[r](8.7,3){$T_\infty^{-1}$}

\psline{->}(13.4,3.4)(13,3.7)
\rput[t](14.2,3.3){Black Droplet}
\rput[b](9.9,4.05){Boundary}
\rput[t](12.5,1.8){(Deformed) Planar BH}
\rput(12.5,0.5){$(b)$}

\end{pspicture}
\caption{A sketch of the relevant solutions: {\bf (a):} black funnel and  {\bf (b):} black droplet above a deformed planar black hole. Both describe possible states of dual field theories in contact with heat baths at temperature $T_\infty$ on spacetimes containing black holes of horizon size $R$. The top line corresponds to the boundary, with the dots denoting the horizon of the boundary black hole.  The shaded regions are those inside the bulk horizons. } \label{f:fundrop}
\end{center}
\end{figure}
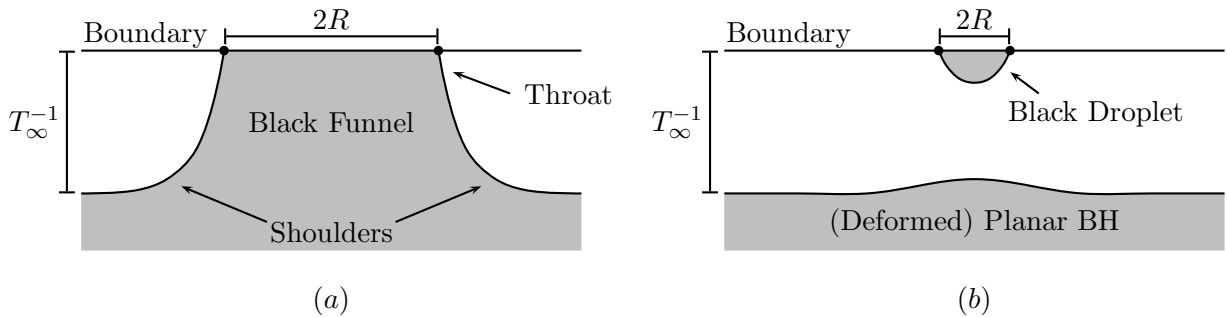

Taking the gauge theory black hole to have radius $R$, the product $RT_\infty$ is scale invariant.  It is thus natural to expect the basic physics to depend crucially on this product, with the precise details of the gauge theory black hole playing a lesser role.  In particular, since depth ($z$) is related to size ($r$) on the boundary, for large $RT_\infty$ one expects funnels to exist and to be thermodynamically dominant, while droplets (above planar black holes) should dominate for small $RT_\infty$.  See \cite{Hubeny:2009ru}  for a more detailed discussion.   It is natural to expect that the two phases are connected by a cone transition as occurs in similar settings \cite{Kol:2002xz,Kol:2003ja,Emparan:2011ve} in which dialing a parameter causes two horizons to meet and perhaps merge.\footnote{This scenario was proposed and explored in \cite{Kol:2002xz} for mergers of horizons with the same temperature.  It was shown in \cite{Emparan:2011ve} that similar behavior can result even when the horizon temperatures differ.  We note that the AdS-Schwarzschild black string solution of \cite{Chamblin:1999by} is such a cone involving a planar black hole of zero temperature.  Since there appears to be no nearby stable solution, the full phase diagram is likely to be as complicated and interesting as that of Kaluza-Klein black holes (see e.g. \cite{Horowitz:2011cq}). }

As in \cite{Hubeny:2009ru}, let us focus for now on the equilibrium context (where $T_{BH} = T_\infty$). In the funnel case, disturbances of the bulk horizon  near $z=0$ (which describes the region near that boundary black hole) can propagate along the bulk  horizon into the region near $r= \infty$ (which describes the thermal plasma).  Although any coherent oscillation is highly damped during this propagation, this merely turns the energy of the oscillation into heat which will nevertheless flow along the horizon.   In contrast, disturbances of the droplet horizon in figure \ref{f:fundrop} (and the heat they generate) cannot directly propagate to the planar black hole horizon.  Instead, they can only couple to the planar black hole horizon via bulk gravity.  Since in linear field theory we describe the coupling between asymptotic scattering states and a black hole in terms of grey body factors, it is natural to say that for the dual field theory the boundary black hole has tiny grey body factors (of order $1/N^2$ or smaller).

Of course, the situation is more complicated than this statement might seem to imply as the dual field theory contains many degrees of freedom that interact strongly.  A more complete story would note that the ${\cal O}(N^0)$ degrees of freedom dual to bulk gravitational waves have more familiar ${\cal O}(1)$ grey body factors while the grey body factors for the $N^2$ degrees of freedom that describe the rest of the plasma near $r = \infty$ appear to vanish exactly to all orders in $1/N$.  It is natural to suppose that these latter grey body factors are in fact exponentially small due to tunneling through a potential barrier of height $N$ or $N^2$.  Presumeably this potential barrier is related to the need to change the ways in which flux tubes connect in attempting to move a quasi-particle from a state in which it is attached via flux tubes to the black hole into a plasma state.  As noted in \cite{Hubeny:2009ru}, the plasma quasi-particles seem to have an effective size which is larger than those of quasi-particles attached to the black hole. Thus one should be able to describe the above potential barrier in terms of an effective potential for the size of a quasi-particle. Again, this is merely a self-consistent interpretation of the bulk gravitational physics.  A complete microscopic understanding in terms of the gauge theory remains to be found.

While it is useful to keep the above general context in mind, the discussion degenerates somewhat in the case of 1+1 CFTs (dual to 2+1 AdS solutions).  In this context, there is no useful definition of the ``size" of a horizon in the CFT.  Indeed, all horizons are analogous to planar horizons in higher dimensions and may therefore be considered to have $R = \infty$.  In particular, there cannot be any effective potential associated with the size of a quasi-particle.  As a result, only the black funnel phase is allowed.

As a final side comment, we mention that it should be possible to construct an even more general set of droplet-phase solutions.  First, note that even in a free field theory, one may consider a thermal state on a black hole background with $T \neq T_{BH}$.  Although the correlation functions are then singular at the horizon, a logical possibility is that that our large $N$ CFT in such a state may be described by a smooth bulk gravitational dual which merely fails to be asymptotically AdS in the usual sense at the horizon of the boundary black hole.  This was the case in \cite{Marolf:2010tg,Hung:2011nu}, which studied an analogous setting involving de Sitter horizons and found that the field theory temperature $T$ corresponded to the temperature of a smooth {\it bulk} horizon that attached to the boundary black hole; i.e., $T = T_{\mathrm{bulk} \ BH} \neq T_{\mathrm{bndy} \ BH}$.  The results of \cite{Marolf:2010tg,Hung:2011nu} thus suggest that general droplet solutions are labeled by three temperatures: $T_{\mathrm{bndy} \ BH}$, $T_{\mathrm{bulk} \ \mathrm{droplet}}$, and $T_\infty$.  In particular, in contrast to the identification in \cite{Figueras:2011va}, even for $T_{BH} \neq 0$ the Boulware (ground) state should be described by a smooth bulk solution having only extreme horizons ($T_{\mathrm{bulk} \ \mathrm{droplet}} = T_\infty = 0$).

\section{Flowing funnels from BTZ black holes}
\label{BTZ}

We now turn to the problem of constructing AdS${}_3$ spacetimes which exhibit heat flow in a stationary state.  Our particular interest concerns bulk solutions dual to a 1+1 CFT on a black hole background.  We refer to such  AdS${}_3$ solutions as  2+1 `flowing funnels.'

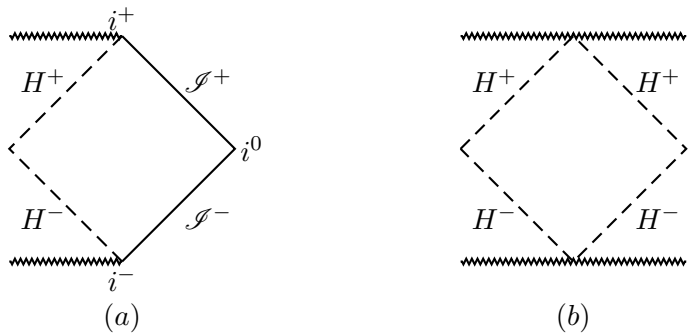
\begin{figure}[h]
\begin{center}
\psset{unit=0.75cm}
\begin{pspicture}(12,6)


\psline[linestyle=dashed](2,1)(0,3)(2,5)
\psline(2,5)(4,3)(2,1)
\pszigzag[coilwidth=0.1,coilarm=0](0,5)(2,5)
\pszigzag[coilwidth=0.1,coilarm=0](0,1)(2,1)

\rput[br](1,4){$H^+$}
\rput[tr](1,2){$H^-$}
\rput[b](2,5.1){$i^+$}
\rput[l](4.1,3){$i^0$}
\rput[t](2,0.9){$i^-$}
\rput[bl](3.1,4){$\mathscr{I}^+$}
\rput[tl](3.1,2){$\mathscr{I}^-$}

\rput(2,0){$(a)$}


\psline[linestyle=dashed](8,3)(10,5)(12,3)(10,1)(8,3)
\pszigzag[coilwidth=0.1,coilarm=0](8,5)(12,5)
\pszigzag[coilwidth=0.1,coilarm=0](8,1)(12,1)

\rput[br](9,4){$H^+$}
\rput[tr](9,2){$H^-$}
\rput[bl](11.1,4){$H^+$}
\rput[tl](11.1,2){$H^-$}

\rput(10,0){$(b)$}

\end{pspicture}
 \caption{Conformal diagrams showing the two types of~$1+1$ background spacetimes for our CFTs. {\bf (a):} A single black hole on the left with an asymptotically flat region on the right.   {\bf (b):} Two black holes.}		
\label{f:bndyBHs}
\end{center}
\end{figure}

Note that in 1+1 dimensions a given static region of spacetime can be attached to no more than two black holes -- one on the left, and one on the right.  We begin with spacetimes that contain only one black hole (say, on the left) and which approach the Minkowski metric in inertial coordinates on the right; see figure~\ref{f:bndyBHs}.  Adding the second black hole will be straightforward once this case is under control.

In fact, let us first consider the case of no black holes at all.  Recall that the stress tensor of stationary CFT states on 1+1 Minkowski space is fully characterized by its right-moving and left-moving temperatures $T_R, T_L$:

\begin{equation}
\label{therm}
ds^2_{CFT} = - du dv, \ \ \ \
T^{CFT}_{ab} dx^a dx^b  = \pi \frac{c}{12} \left(T_R^2 du^2 + T_L^2 dv^2\right),
\end{equation}
where we have introduced null coordinates $u = t-x, v= t+x$ , and $c$ is the central charge of the CFT.    The system may also be characterized by a temperature $T = 2(T_L^{-1} + T_R^{-1})^{-1}$ and a chemical potential for momentum $\mu =  \ell (T_R - T_L)/(T_R + T_L)$.  Although 1+1 CFTs are not well-described by perfect fluids, one may nevertheless think of the system as being `at rest' when $T_L = T_R$, as there is then no net transport of energy.  Boosting to a more general frame, it is natural to define the `velocity' of the system to be $U^a\partial_a  = \frac{1}{\sqrt{T_L T_R}} \left( T_R \partial_v + T_L \partial_u \right)$.

As is well-known, the bulk spacetimes dual to such flowing thermal states are just AdS${}_3$ in BTZ coordinates \cite{Banados:1992wn,Banados:1992gq}; i.e., they are BTZ black holes with the `angle' unwrapped so that it runs over $(-\infty, \infty)$ instead of over $S^1$.  In the present context, it is natural to give the would-be BTZ angle the name $x$ and units of length.  The bulk metric then takes the form

\begin{equation}
\label{BTZmet}
ds^2_\mathrm{bulk} = -\frac{(\rho^2-\rho_+^2)(\rho^2-\rho_-^2)}{\ell^2 \rho^2} \, dt^2 + \frac{\rho^2\ell^2 }{(\rho^2-\rho_+^2)(\rho^2-\rho_-^2)} \, d\rho^2 + \frac{\rho^2}{\ell^2} \left(dx - \frac{\rho_+ \rho_-}{\rho^2}\, dt\right)^2,
\end{equation}
where $\ell$ is the usual AdS scale and the temperatures are~$T_R = (\rho_+ + \rho_-)/2\pi\ell$ and~$T_L = (\rho_+ - \rho_-)/2\pi\ell$. We take $\rho_+ > 0$ so that the sign of $\rho_-$ determines the sign of the BTZ angular momentum $J$.  In \eqref{BTZmet}, we have called the usual BTZ radial coordinate $\rho$ in order to reserve $r$ for a radial coordinate along the boundary. The boundary stress tensor \cite{Henningson:1998gx,Balasubramanian:1999re} of
\eqref{BTZmet} is \eqref{therm} with $c = 3\ell/2G$, where $G$ is the bulk gravitational constant.

Now, a general static 1+1 spacetime may be written as $ds^2 = \Omega^2(x) \left(-dt^2 + dx^2\right)$, and so may be generated from Minkowski space via an appropriate conformal rescaling.  In particular, we obtain a black hole of temperature $T_{\mathrm{bndy} \ BH} = \kappa/2 \pi$ by taking $\Omega \rightarrow 1$ as $x \rightarrow +\infty$ and $\Omega \sim e^{\kappa x}$ as $x \rightarrow - \infty$. Under such a conformal rescaling, the CFT stress tensor transforms as~\cite{DiFrancesco:1997} 
\begin{equation}
T_{ab} \rightarrow T_{ab} + \frac{c}{12 \pi }\left[\nabla_a \nabla_b \sigma - \nabla_a \sigma \nabla_b \sigma + \frac{1}{2} g_{ab} \left(\nabla \sigma\right)^2 - g_{ab} \nabla^2 \sigma \right]
\end{equation}
where $\Omega = e^\sigma.$  In particular, the stress tensor is unchanged at large positive $x$ (where $\nabla_a \sigma$ vanishes) while at large negative $x$ eqn. \eqref{therm} becomes
\begin{equation}
T_{CFT} = \pi \frac{c}{12} \left[ e^{2 \kappa u} \left(T_R^2 - \frac{1}{4\pi^2} \kappa^2  \right) dU^2 + e^{-2 \kappa v} \left(T_L^2 - \frac{1}{4\pi^2} \kappa^2 \right) dV^2 \right] + \dots
\end{equation}
in terms of the new affinely-parametrized null coordinates $U,V$ (which satisfy
$U = -\kappa^{-1} e^{-\kappa u} + \dots$ and $V = \kappa^{-1} e^{\kappa v} + \dots$ where $+ \dots$ represents terms that are subleading as $x \rightarrow -\infty$).   Note that the new stress tensor is regular on the future  horizon ($u = \infty$)  precisely when we choose $T_R = T_{\mathrm{bndy} \ BH}$. 

For this choice, one may interpret the CFT state as describing heat exchange between a boundary black hole of temperature $T_{\mathrm{bndy} \ BH} = T_R$ and a heat bath at infinity ($x = + \infty)$ with temperature $T_L$.  In particular, suppose that we instead choose the background metric to be time dependent, and to evolve from 1+1 Minkowski space in the far past to the desired boundary black hole spacetime in the far future.  Then, by the usual arguments, when the CFT begins in an initial thermal state of temperature $T_L$ the late-time behavior of the CFT is described by our solution above.\footnote{Recall that $T_L$ is the temperature of the left-moving part of the CFT.  Unfortunately, this is naturally thought of as being sourced by a heat bath on the right (at $x = +\infty$).}

Returning to the stationary case,
let us remark on several features of our bulk solution:

\begin{itemize}

\item The choice $T_L=0$ is dual to the Unruh state of the CFT.  In particular, the Unruh state corresponds to an extreme horizon in the bulk whose properties are fixed by the temperature of the boundary black hole.
In contrast, note that the Boulware state is described by an `unwrapped' $M=0$ BTZ black hole with $T_L=T_R=0 \neq T_{\mathrm{bndy} \ BH}$, which is just a Poincar\'e horizon (independent of $T_{\mathrm{bndy} \ BH}$).  As noted in section \ref{review}, this is a smooth bulk spacetime with a singular boundary stress tensor.  In this case, the singular boundary stress tensor is due to a singular choice of conformal frame.

\item By construction, on the boundary of our spacetime the stationary Killing field $\partial_t$ of BTZ agrees with the static Killing field of the boundary metric.  But BTZ has a second (commuting) Killing field $\partial_x$.  While this is not a Killing field of the boundary black hole, the fact that all 1+1 metrics are conformally flat means that $\partial_x$  acts as a  conformal Killing field on the boundary.  This is a peculiar feature of our AdS${}_3$ problem that will not be reflected in higher dimensions.  Indeed, in higher dimensions it is easy to show that any conformal isometries of boundary metrics describing  non-extreme stationary spherically symmetric boundary black holes are in fact boundary Killing fields, so that there can be no such `accidental' Killing fields in the bulk.\footnote{Extreme black holes can have such conformal isometries but do not by themselves lead to heat flow.}

\item The bulk solution has a Killing horizon generated by $\chi = \partial_t + \ell^{-1} \mu \partial_x$, where $\mu = \ell(T_R -T_L)/(T_R + T_L)$ as above is related to the angular velocity of the horizon via $\mu = \Omega_H \ell$.  Thus we see that $\mu$ characterizes the rate at which null generators of the horizon pass from one Killing orbit to another.   For $\mu > 0$ we may say that, in this sense, generators `move' from the boundary down into the bulk and toward positive $x$, while for $\mu < 0$ the `motion' is toward negative $x$ and up toward the boundary.  As was the case for $\partial_x$ above, $\chi$ is an accidental symmetry from the viewpoint of the boundary theory and will have no analogue in higher dimensions; i.e., the black holes that describe flowing funnels dual to $d$ $>$ 1+1 CFTs will not have Killing horizons.

\item For $T_L = T_R$ our solution is precisely the static AdS${}_3$ black funnel constructed in \cite{Hubeny:2009ru}.
\end{itemize}

In many implementations of the AdS/CFT correspondence one may gain insight into the CFT state by displaying the bulk solution in Fefferman-Graham coordinates.   This seems to be less useful in the current context as these coordinates are highly singular.  We relegate the details of the coordinate transformations and the resulting metrics to the appendix, though we briefly summarize the key points below.

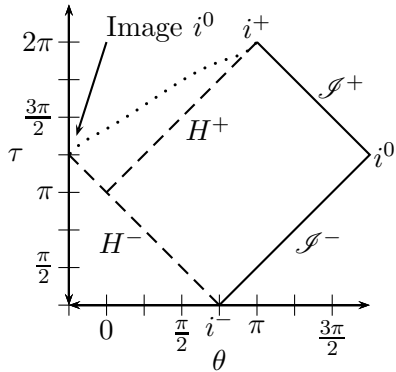
\begin{figure}[h]
\begin{center}
\psset{unit=0.5cm}
\begin{pspicture}(-3,-1)(8,10)

\psaxes[labels=none,Dx=1]{<->}(-1,1)(-1,1)(7,9)
\rput[r](-1.4,8){$2\pi$}
\rput[r](-1.4,6){$\frac{3\pi}{2}$}
\rput[r](-1.4,4){$\pi$}
\rput[r](-1.4,2){$\frac{\pi}{2}$}
\rput[t](6,0.6){$\frac{3\pi}{2}$}
\rput[t](4,0.6){$\pi$}
\rput[t](2,0.6){$\frac{\pi}{2}$}
\rput[t](0,0.6){0}
\rput[r](-2.2,5){$\tau$}
\rput[t](3,-0.2){$\theta$}

\psline[linestyle=dashed](3,1)(-1,5)
\psline[linestyle=dashed](0,4)(4,8)
\psline(3,1)(7,5)(4,8)
\fileplot[linestyle=dotted,linewidth=1.2pt]{GlobalData.dat}

\rput[tr](1,3){$H^-$}
\rput[tl](2.1,6){$H^+$}
\rput[tl](5,3){$\mathscr{I}^-$}
\rput[bl](5.6,6.5){$\mathscr{I}^+$}
\rput[l](7.1,5){$i^0$}
\rput[b](4,8.1){$i^+$}
\rput[t](3,0.8){$i^-$}
\psline{->}(0,8)(-0.8,5.5)
\rput[bl](0,8){Image~$i^0$}

\end{pspicture}
 \caption{The relevant portion of the AdS$_3$ boundary in global coordinates. The dotted line indicates the singularity of the boundary metric \eqref{bndyBH}; i.e., a singularity in the conformal frame associated with \eqref{bndyBH}.  The boundary past horizon $H^-$ is another such singularity.  Since $\theta$ has period $2\pi$, the point marked ``Image $i^0$'' represents the same event on the boundary of global AdS$_3$ as does $i^0$. Due to singularities in the change of conformal frame it nevertheless represents a distinct point of the black hole boundary spacetime \eqref{bndyBH}.}
\label{f:bndy}
\end{center}
\end{figure}

\begin{figure}[t!]
\psset{unit=0.8cm}
\begin{center}
\begin{pspicture}(0,0)(20.4,12)

\rput(5,6){\includegraphics[width=0.5\textwidth,angle=90]{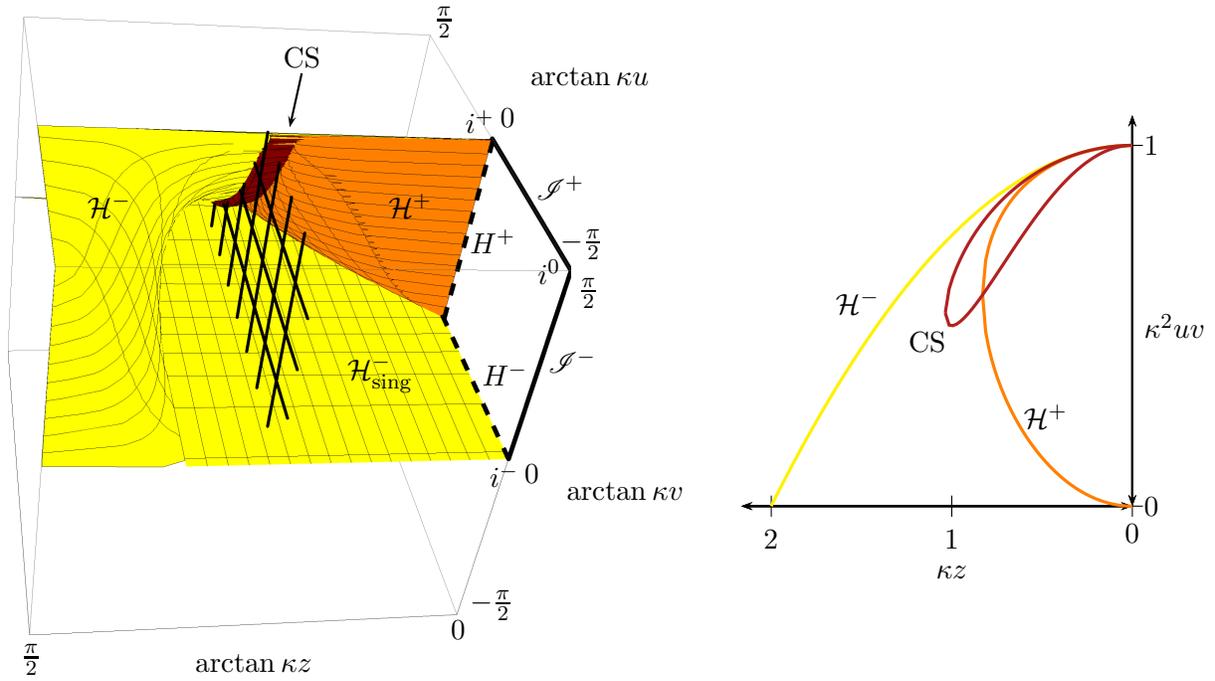}}

\rput[bl](9.5,7.1){$-\frac{\pi}{2}$}
\rput[bl](8.5,9.3){$0$}
\rput[bl](7.4,10.8){$\frac{\pi}{2}$}
\rput[bl](9,10){$\arctan\kappa u$}

\rput[t](0.7,0.7){$\frac{\pi}{2}$}
\rput[t](7.8,1.1){$0$}
\rput(4.4,0.4){$\arctan \kappa z$}

\rput[tl](8,1.6){$-\frac{\pi}{2}$}
\rput[tl](8.9,3.7){$0$}
\rput[tl](9.8,6.8){$\frac{\pi}{2}$}
\rput[tl](9.6,3.4){$\arctan \kappa v$}

\rput[bl](9.1,8.1){$\mathscr{I^+}$}
\rput[tl](9.3,5.6){$\mathscr{I^-}$}
\rput[tl](8,7.6){$H^+$}
\rput[bl](8.2,5){$H^-$}
\rput(7,8){$\mathcal{H}^+$}
\rput(2,8){$\mathcal{H}^-$}
\rput[br](8.4,9.2){$i^+$}
\rput[tr](8.8,3.7){$i^-$}
\rput[r](9.5,6.9){$i^0$}
\psline{->}(5.2,10.2)(5,9.3)
\rput[b](5.2,10.3){CS}
\rput(6.5,5.2){$\mathcal{H}_\mathrm{sing}^-$}


\psaxes[Dx=3,Dy=6,labels=none]{<->}(19,3)(12.5,3)(19,9.5)

\def\z{x 19 sub 3 div}
\psplot[linewidth=1.2pt,plotpoints=50,linecolor=orange]{16.5147187}{19}{4 x 19 sub 3 div 2 exp add 16  24 x 19 sub 3 div 2 exp mul sub x 19 sub 3 div 4 exp add sqrt add 8 div 6 mul 3 add} 
\psplot[linewidth=1.2pt,plotpoints=50,linecolor=orange]{16.5147187}{19}{4 x 19 sub 3 div 2 exp add 16  24 x 19 sub 3 div 2 exp mul sub x 19 sub 3 div 4 exp add sqrt sub 8 div 6 mul 3 add} 
\psplot[linewidth=1.2pt,plotpoints=50,linecolor=yellow]{13}{19}{4 x 19 sub 3 div 2 exp sub 4 div 6 mul 3 add} 
\psplot[linewidth=1.2pt,plotpoints=50,linecolor=darkred]{15.8941716}{19}{16 8 x 19 sub 3 div 2 exp mul sub x 19 sub 3 div 4 exp add 16 x 19 sub 3 div 4 exp mul 16 x 19 sub 3 div 6 exp mul sub x 19 sub 3 div 8 exp add sqrt add 16 div 6 mul 3 add} 
\psplot[linewidth=1.2pt,plotpoints=50,linecolor=darkred]{15.8941716}{19}{16 8 x 19 sub 3 div 2 exp mul sub x 19 sub 3 div 4 exp add 16 x 19 sub 3 div 4 exp mul 16 x 19 sub 3 div 6 exp mul sub x 19 sub 3 div 8 exp add sqrt sub 16 div 6 mul 3 add} 

\rput[l](19.2,9){1}
\rput[l](19.2,3){0}
\rput[t](19,2.7){0}
\rput[t](16,2.6){1}
\rput[t](13,2.6){2}
\rput[l](19.2,6){$\kappa^2 uv$}
\rput[t](16,2){$\kappa z$}

\rput[bl](17.2,4.3){$\mathcal{H}^+$}
\rput[br](14.8,6.2){$\mathcal{H}^-$}
\rput[tr](15.9,5.9){CS}

\end{pspicture}

\caption{{\bf Left:} Notable features of the bulk AdS$_3$ spacetime in Fefferman-Graham (FG) coordinates $u,v,z$, (compactified to show their full range) for our Unruh state solution ($T_L =0, \kappa = 2\pi T_R$) and boundary metric \eqref{bndyBH}.  Past and future bulk horizons~$\mathcal{H}^\pm$, boundary horizons~$H^\pm$ and the various pieces $\mathscr{I}^\pm$,~$i^{0,\pm}$ of infinity for the boundary spacetime (see fig. \ref{f:bndy}) are shown.   The surfaces ${\cal H}^-_\mathrm{sing}$ (at $v=0$) and CS are FG coordinate singularities.  The first
(${\cal H}^-_\mathrm{sing}$) acts like part of the bulk past horizon from the FG point of view, though its image in global coordinates coincides with that of $i^-$.
The second (CS) is a closed surface from the FG point of view which begins and ends at $z=0, uv=1$ (the singularity of the boundary black hole; dotted line in figure \ref{f:bndy}).  Parts of the surfaces CS, ${\cal H}^\pm$ are obscured, as is a third FG coordinate singularity at  $\kappa^2 uv=1$ (for all $z$).   A line on the CS surface near its maximum value of $z$ is a set of branch points.  The black grid on the left diagram marks the image of an associated branch cut chosen so that the transformation to FG coordinates is one-to-one in the region bounded by this cut, the CS surface, and the plane $z=0$.  {\bf Right:} The general structure and the relative locations of CS, ${\cal H}^\pm$ are made clear. These surfaces depend only on the product $uv$.  CS and ${\cal H}^+$ intersect along a single line.   ${\cal H}^+$ and ${\cal H}^-$ do not intersect in the bulk but meet  only at the singularity of the boundary black hole.  }
\label{f:FG}
\end{center}
\end{figure}

\begin{figure}[h!]
\begin{center}
\psset{unit=0.8cm}
\begin{pspicture}(0,0)(14,13)

\rput(3,6.5){\includegraphics[width=0.3\textwidth]{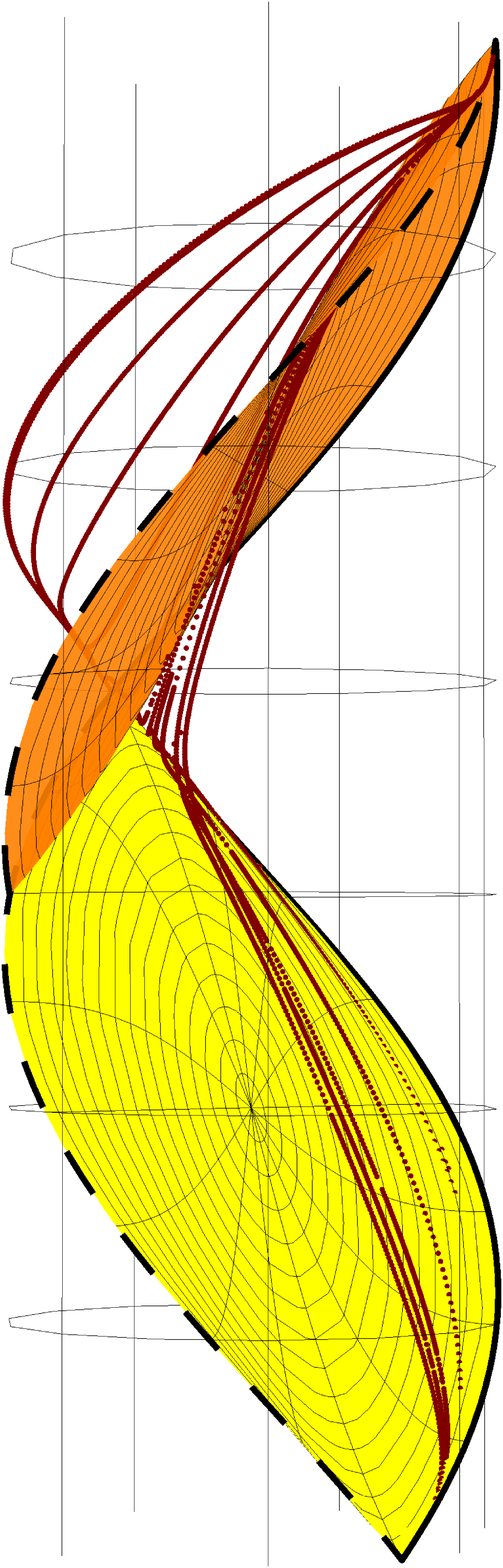}}
\rput(10,6.3){\includegraphics[width=0.3\textwidth]{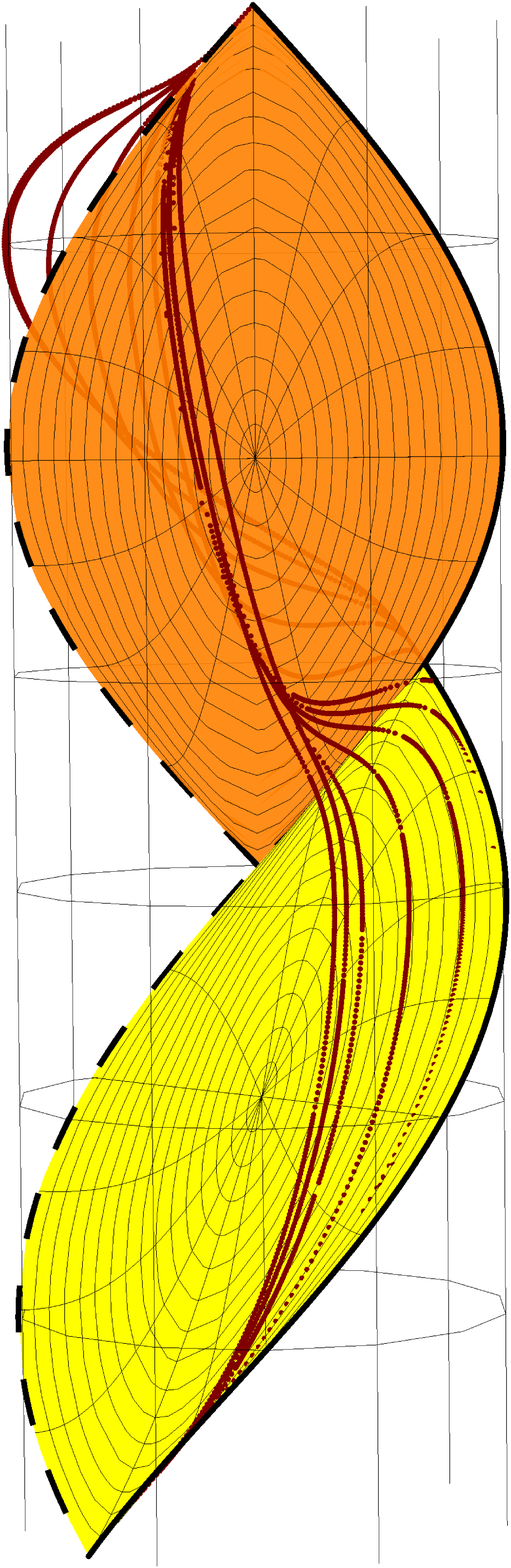}}


\rput[r](1.1,12.2){$2\pi$}
\rput[r](1.1,8.8){$\frac{3\pi}{2}$}
\rput[r](1.1,5.6){$\pi$}
\rput[r](1.1,2.4){$\frac{\pi}{2}$}
\rput(0,7){$\tau$}

\rput[l](12.1,12.2){$2\pi$}
\rput[l](12.1,8.8){$\frac{3\pi}{2}$}
\rput[l](12.1,5.6){$\pi$}
\rput[l](12.1,2.3){$\frac{\pi}{2}$}
\rput(13,7){$\tau$}

\rput[t](4.2,0.5){$i^-$}
\rput[b](5,12.2){$i^+$}
\rput[tr](2.4,2.6){$H^-$}
\rput[br](1.4,7){$H^+$}
\rput(3,4){$\mathcal{H}^-$}
\rput(2,7.2){$\mathcal{H}^+$}
\rput[tl](4.4,10){$\mathscr{I}^+$}
\rput[bl](4.7,4){$\mathscr{I}^-$}
\rput(2.2,9.6){CS}

\rput[t](8.7,0.4){$i^-$}
\rput[b](10,12.4){$i^+$}
\rput[l](11.4,7.3){$i^0$}
\rput[br](8.7,4){$H^-$}
\rput[tr](8.2,8){$H^+$}
\rput(10.5,9){$\mathcal{H}^+$}
\rput(9.5,3.6){$\mathcal{H}^-$}
\rput[bl](11.4,10.6){$\mathscr{I}^+$}
\rput[tl](10.8,3){$\mathscr{I}^-$}
\rput(10.6,5.5){CS}

\end{pspicture}
 \caption{Notable features of the bulk~AdS$_3$ spacetime in global coordinates associated with the Unruh state  ($T_L =0, \kappa = 2\pi T_R$) and the boundary black hole \eqref{bndyBH}. The bulk past and future horizons ${\cal H}^\pm$ are labeled, as are  $H^\pm$, $\mathscr{I}^{\pm}$ and~$i^{0,\pm}$ describing the horizons and infinity of the boundary metric  (see figure \ref{f:bndy}). In global coordinates, the surface ${\cal H}^-_\mathrm{sing}$ of figure \ref{f:FG} coincides with $i^-$.
    The lines labeled CS are~$uv = \mathrm{const.}$ contours of the FG coordinate singularity described by the CS surface of figure \ref{f:FG}.  In Fefferman-Graham coordinates, the CS surface is closed and pinches off as it reaches the boundary singularity at~$\kappa^2 uv = 1$.  In global coordinates, this behavior gives rise to an open surface with two edges: one edge is at the boundary singularity  (seen above behind the bulk horizon~$\mathcal{H}^+$; this line is $\kappa^2 uv=1$ for all values of the FG $z$ coordinate) and $\mathscr{I}^-$.  The other edge travels from~$i^+$ to~$i^-$ inside the bulk.  The associated branch cut shown as a black grid in figure \ref{f:FG} is not drawn, but would start near this edge in the bulk and fold back to the right, passing above the CS surface and terminating on $\mathscr{I}^+$ and $\mathscr{I}^-$. One should be aware that the part of the CS surface above the branch cut (near the interior edge) has multiple images in figure \ref{f:bndy} and is a coordinate singularity only after one has passed through the branch cut.}		
\label{f:global}
\end{center}
\end{figure}

There are three natural sources of coordinate singularities in the Fefferman-Graham coordinates associated with any boundary black hole: i) the past horizon $H^-$ of the boundary black hole, ii) null infinity $\mathscr{I}^\pm$ of the boundary spacetime (see figure \ref{f:bndyBHs}), and iii)  the singularity of the boundary black hole.  The singularity on $H^-$ is associated with the fact that, while we tuned parameters to make the CFT stress tensor smooth across the future horizons, it  generally remains singular on $H^-$. The problem at $\mathscr{I}^\pm$ is associated with the fact that these are finite locations when AdS${}_3$ is described in global coordinates\footnote{\label{global} The particular global coordinates used in figures \ref{f:bndy}, \ref{f:FG}, and \ref{f:global} below are the dimensionless $\tau,R,\theta$ for which
$$
ds^2 = \frac{4\ell^2}{(1-R^2)^2} \left[- \frac{1}{4}(1+R^2)^2 d\tau^2 + dR^2 + R^2 d\theta^2 \right] .
$$}.  In addition, the singularity of the boundary black hole is clearly a singularity of the transformation between any global coordinates and our Fefferman-Graham coordinates.  Since the boundary metric can be chosen at will (and need not satisfy any equations of motion) we are free to place this singularity anywhere we like inside the boundary horizon.  But it is important to note that, when written in conformally flat form, the conformal factor $\Omega$ of any boundary black hole metric necessarily has some singular feature associated with the fact that e.g. future-directed null geodesics along $\mathscr{I}^+$ encounter the horizon $H^+$ only at infinite affine parameter while other left-moving future-directed null geodesics encounter the horizon at finite affine parameter.

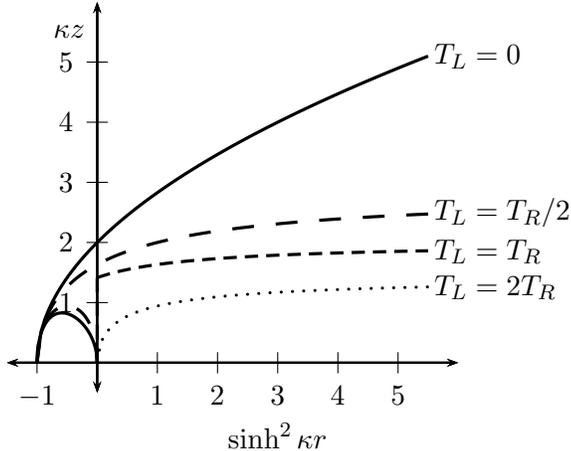
\begin{figure}[h!]
\begin{center}
\psset{unit=.8cm}
\begin{pspicture}(-1.5,-1)(7,6)

\psaxes[Dx=1]{<->}(0,0)(-1.5,-0.5)(6,6)

\psplot[plotpoints=100, linestyle=dotted, linewidth=1.2pt]{0}{5.5}{x 1 x add mul 1 2 x mul add 2 x add mul div sqrt 2 mul}
\rput[l](5.6,1.25){$T_L = 2T_R$}

\psplot[plotpoints=100, linestyle=dashed, linewidth=1.2pt]{0}{5.5}{1 x add 2 x add div sqrt 2 mul}
\psline[linestyle=dashed,linewidth=1.2pt](0,0)(0,1.41)
\rput[l](5.6,1.85){$T_L = T_R$}

\psplot[plotpoints=100,linestyle=dashed, dash=8pt 8pt, linewidth=1.2pt]{-1}{0}{2 x mul 1 x add mul neg 1 x sub  2 x add mul div sqrt 2 mul}
\psplot[plotpoints=100,linestyle=dashed, dash=8pt 8pt, linewidth=1.2pt]{-1}{5.5}{2 1 x add mul 3 x add div sqrt 2 mul}
\rput[l](5.6,2.5){$T_L = T_R/2$}

\psplot[plotpoints=100,linewidth=1.2pt]{-1}{0}{x 1 x add mul neg 2 x add div sqrt 2 mul}
\psplot[plotpoints=100,linewidth=1.2pt]{-1}{5.5}{1 x add sqrt 2 mul}
\rput[l](5.6,5.1){$T_L = 0$}

\rput[r](-0.2,5.5){$\kappa z$}
\rput[t](3,-1){$\sinh^2 \kappa r$}

\end{pspicture}

\caption{The future horizon in Fefferman-Graham coordinates for  $T_L/T_R = 2,1,\frac{1}{2},0$ and $\kappa = 2\pi T_R$ (smooth future horizon).  The case~$T_L = T_R$ describes the Hartle-Hawking state studied in~\cite{Hubeny:2009ru}. The case~$T_L = 0$ describes the Unruh case studied in detail in this paper.  The horizontal axis plots $\sinh^2 \kappa r$ rather than just~$r$ in order to reach all the way to the boundary singularity at~$\sinh^2 \kappa r = -1$.  The curve diverges at large $r$ for $T_L =0$ but otherwise asymptotes to $\kappa z = 2\sqrt{T_R/T_L}$. }		
\label{f:horizons}
\end{center}
\end{figure}

It is no surprise that the full singularities of the Fefferman-Graham coordinate system extend into the bulk, connecting to the boundary at the above three locations.  One might hope that the singularities remain localized at the the bulk BTZ singularity and at the natural bulk null surfaces associated with (i) and (ii) above.  But that turns out not to be the case, and Fefferman-Graham coordinate singularities extend outside the horizons of the BTZ black hole. The situation is summarized in figures \ref{f:bndy}, \ref{f:global}, and \ref{f:FG} below for the AdS{}$_3$ spacetime dual to the Unruh state of the CFT on the metric
\begin{equation}
\label{bndyBH}
ds^2_{CFT} = - \tanh^2 \kappa r \, dt^2 + dr^2 = \frac{-1}{1-\kappa^2 uv} \, du dv,
\end{equation}
where $\kappa=2\pi T_R$ is again the surface gravity of the boundary black hole.  The general case $T_L \neq 0$ is similar. We draw the reader's attention to the branch cut in figures \ref{f:global} and \ref{f:FG}, which limits the utility of Fefferman-Graham coordinates to a region surprisingly close to the boundary.  We also include plots of the Fefferman-Graham $z$ vs. $r$ along the future horizons for various values of $T_L/T_R$ (see figure \ref{f:horizons}).

Finally, let us consider the addition of a 2nd black hole to the 1+1 CFT spacetime. This may be accomplished by performing another conformal rescaling, this time with $\Omega \sim e^{-\kappa_R x}$ at large positive $x$.  It is natural to choose $\Omega$ so that $|\partial_t|^2 = -1$ at e.g. $x=0$.  This provides a preferred location with respect to which to normalize notions of temperature and surface gravity. We also rename the above $\kappa$ as $\kappa_L$ to refer to the surface gravity of the left black hole.  The resulting family of solutions is labeled by four parameters $(T_L, T_R, \kappa_L, \kappa_R)$, and the CFT stress tensor is smooth across both horizons if $T_R = \kappa_L/2\pi$ and $T_L = \kappa_R/2\pi$.  Such states describe the asymptotic future of {\it any} CFT state (with a smooth stress tensor) on a spacetime which evolves from flat Minkowski space in the far past to one containing our two black holes in the far future.  The details are in direct analogy to the case of a single black hole above.

\section{Discussion}
\label{disc}

We showed above how AdS${}_3$ solutions dual to stationary CFT${}_2$ states with heat transport between black holes (or between one black hole and a heat bath at infinity) can be constructed by `unwrapping' the angular coordinate of BTZ black holes and changing conformal frame at infinity.  Thus the solutions may also be described as pure AdS${}_3$ in an appropriate conformal frame.  An interesting point was that Unruh states of the CFT (living on a spacetime with a single black hole) are dual to extreme horizons in the bulk.  But perhaps the notable feature of our solutions is just that they are everywhere smooth, despite the fact that the boundary stress tensor is generally singular on the past horizon of the boundary black hole.
This smoothness is similar to the phenomenon seen in \cite{Marolf:2010tg,Hung:2011nu} for CFTs on black hole spacetimes in which the CFT was forced to have a temperature different from the Hawking temperature of the boundary black hole ($T_{CFT} \neq T_{\mathrm{bndy} \ BH}$). See also \cite{Headrick:2010zt}.

In fact, ref. \cite{Marolf:2010tg,Hung:2011nu} considered AdS${}_{d+1}$ solutions for all $d$ $\ge$ 1+1 and found smooth bulk solutions in all dimensions.   But as we now discuss it is difficult to see how this could be possible for higher dimensional flowing funnels.  In the AdS${}_3$ case, the `accidental' Killing fields discussed in section \ref{BTZ} meant that our solutions had Killing horizons.  As a result, the points along a given null generator were related by a symmetry and nothing interesting can happen as one follows a null generator into the distant past.  But section \ref{BTZ} noted that there are no accidental symmetries in the higher-dimensional case.  Furthermore, as in our AdS${}_3$ case, the flow of heat should make the stationary Killing field $\partial_t$ fail to be null on the horizon\footnote{At least near the boundary, one may think of this as due to the fact that non-zero $tr$ components of the boundary stress tensor force $\partial_t$ to be non-hypersurface-othogonal, and thus to differ from the horizon-generating vector field.}.  Thus the generators will again move with respect to the stationary Killing orbits.

To understand the implications, let us consider a flowing AdS${}_{d+1}$ funnel with SO($d-1$) rotational symmetry.
Here it is useful to use the size $r$ of the $S^{d-2}$ spheres of symmetry as a coordinate along the horizon.
Now, the twist $\omega_{ab}$ of the horizon-generating vector field $\xi^a$ is an anti-symmetric tensor on the horizon satisfying $\xi^a \omega_{ab} =0 $; i.e., it effectively has components only in the $r$-direction and in the angular direction.  So by the SO($d-1$) symmetry $\omega_{ab} =0$ and the null generators necessarily focus at finite affine parameter if the expansion or shear is non-zero.

Note that the value of $r$ will change as one moves along a null generator of the horizon.  This means that one of the eigenvalues of $\nabla_{(a} \xi_{b)}$ must be non-zero and thus that the expansion and shear cannot both vanish.  Since horizon generators necessarily extend to infinite affine parameter toward their future, we conclude that they focus at finite affine parameter in the past.

There are now two logical possibilities: 1) that this focusing represents a caustic in a smooth spacetime, or 2) that it represents a spacetime singularity.  While we have not completely ruled out the first option, the 2nd seems much more natural.  This is just the picture suggested originally in \cite{Hubeny:2009ru}.  It remains an interesting challenge to construct such higher-dimensional flowing funnel solutions.

\subsection*{Acknowledgements}
DM thanks Veronika Hubeny and Mukund Rangamani for many interesting discussions of flowing funnels over several years and Ted Jacobson for discussions concerning horizons with varying temperatures.  He also thanks the
Centro de Ciencias de Benasque Pedro Pascual for its hospitality during the July 2011 Gravity Workshop in July 2011.  Finally, he thanks the participants of that workshop, and especially Roberto Emparan, Pau Figueras, Nemaja Kaloper, Rob Myers, and Toby Wiseman, for many stimulating conversations.  This work was supported in part by the US National Science Foundation under grant PHY08-55415 and by funds from the University of California.

\appendix

\section{Flowing Funnels in Fefferman-Graham coordinates}

The goal of this appendix is to describe the solutions of section \ref{BTZ} in Fefferman-Graham coordinates associated with the boundary metric \eqref{bndyBH}.  This boils down to computing the relevant coordinate transformation between these new coordinates and e.g. the BTZ coordinates $ t, \rho, x$ of \eqref{BTZmet}. We will in fact use null Fefferman-Graham coordinates  $u,v,z$ with
$u = -\kappa^{-1} e^{-\kappa t} \sinh \kappa r$, $v = \kappa^{-1} e^{\kappa t} \sinh \kappa r$ so that (using the Fefferman-Graham gauge conditions) the bulk metric takes the form
\begin{equation}
\label{FGmetric}
ds^2 = \frac{\ell^2}{z^2}\left[-\left(\frac{1}{1-\kappa^2 uv} + \mathcal{O}(\kappa^2 z^2)\right) \, du \, dv + \mathcal{O}(\kappa^2 z^2) \, du^2 + \mathcal{O}(\kappa^2 z^2) \, dv^2 + dz^2\right].
\end{equation}

While it is in principle possible to compute this transformation directly, we find it simpler to use Poincar\'e coordinates $U_P,V_P,Z_P$ on AdS${}_3$, for which the AdS${}_3$ metric is
$ds^2 = \ell^2(-dU_P \, dV_P + dZ^2_P)/Z^2_P$,
as an intermediate step.  The point here is that the transformation from ($\rho,t,x$) to ($U_P,V_P,Z_P$) is known explicitly.  We may then change conformal frame and solve for  ($u,v,z$) in terms of ($U_P,V_P,Z_P$) as a power series in $z$ and,  due to the relative simplicity of the Poincar\'e patch metric, the result is a geometric series that is easily summed and written in closed form.  Combining the two transformations then gives the desired result.  In contrast, summing the series solution to go directly from ($\rho,t,x$) to ($u,v,z$) is more difficult.

Let us look for Fefferman-Graham coordinates in which the boundary metric takes the form \eqref{FGmetric} by making the ansatz that $U_P,V_P,Z_P$ have a series expansion in (integer) powers of $z$ and that the leading ${\cal O}(z^0)$ terms in $U_P,V_P$ are respectively $\ell (\kappa u)^{\gamma_R}$, $\ell (\kappa v)^{\gamma_L}$, while the leading term in $Z_P$ is
$\ell \sqrt{\gamma_R \gamma_L (\kappa u)^{\gamma_L-1} (\kappa v)^{\gamma_R-1}(1-\kappa^2 uv)} \, \kappa z$.
Then as stated above one finds that the result is a geometric series.  Summing the series yields
\begin{eqnarray}
\label{PtoFG}
U_P(u,v,z) &=& \ell (\kappa u)^{\gamma_R}\left[1 - 2\gamma_R \frac{(1-\kappa^2 uv)[1-\gamma_L(1-\kappa^2 uv)](\kappa z)^2}{4\kappa^2 uv(1-\kappa^2 uv) - [1-\gamma_L(1-\kappa^2 uv)][1-\gamma_R(1-\kappa^2 uv)](\kappa z)^2}\right], \cr
V_P(u,v,z) &=& \ell (\kappa v)^{\gamma_L}\left[1 - 2\gamma_L\frac{(1-\kappa^2 uv)[1-\gamma_R(1-\kappa^2 uv)](\kappa z)^2}{4\kappa^2 uv(1-\kappa^2 uv) - [1-\gamma_L(1-\kappa^2 uv)][1-\gamma_R(1-\kappa^2 uv)](\kappa z)^2}\right], \cr
Z_P(u,v,z) &=& \ell \sqrt{\gamma_L\gamma_R} \frac{4(\kappa u)^{(\gamma_L+1)/2} (\kappa v)^{(\gamma_R+1)/2}(1-\kappa^2 uv)^{3/2}\kappa z}{4\kappa^2 uv(1-\kappa^2 uv) - [1-\gamma_L(1-\kappa^2 uv)][1-\gamma_R(1-\kappa^2 uv)](\kappa z)^2}.
\end{eqnarray}
It is then straightforward to read off the boundary stress tensor \cite{Henningson:1998gx,Balasubramanian:1999re} to obtain
\begin{align}
\label{st}
T_{uu} &= \frac{c}{12\pi}\kappa^2 \left[\frac{\kappa^2 v^2}{4(1-\kappa^2 uv)^2} + \frac{\gamma_R^2 - 1}{4\kappa^2 u^2}\right], \cr
T_{vv} &= \frac{c}{12\pi}\kappa^2 \left[\frac{\kappa^2 u^2}{4(1-\kappa^2 uv)^2} + \frac{\gamma_L^2 - 1}{4\kappa^2 v^2}\right], \cr
T_{uv} &= -\frac{c}{12\pi} \, \frac{\kappa^2}{2(1-\kappa^2 uv)^2},
\end{align}
where as usual $c = 3\ell/2G.$  Note that regularity of $T_{ab}$ on the future horizon ($u=0$) would require $\gamma_R =1$, while a vanishing incoming flux at $\mathscr{I}^-$ ($u = -\infty$) would require $\gamma_L =0$.  Thus the transformations \eqref{PtoFG} degenerate in the Unruh state, as they suggest that $V_P$ is independent of $(u,v,z)$.

This subtlety disappears when one instead transforms to BTZ coordinates.
The transformation from $(U_p,V_p,Z_p)$ to $(t,\rho,x)$ is given for~$\rho > \rho_+$ by equations~(2.9) of~\cite{Carlip:1995qv} and takes a simple form in terms of the null Fefferman-Graham coordinates $U = -\ell e^{-\kappa(t-x)}$, $V = \ell e^{\kappa(t+x)}$, and $Z$ given by the implicit relation
\begin{equation}
 \rho = \frac{\ell^2}{Z}\sqrt{1+ \left(\Delta^2 + \Sigma^2\right) (Z/\ell)^2 + \Delta^2\Sigma^2  (Z/\ell)^4}, \end{equation}
for $\Delta = (\rho_+ - \rho_-)/2\ell$ and $\Sigma = (\rho_+ - \rho_-)/2\ell$. We mention that the metric in $(U,V,Z)$ coordinates is
\begin{equation}
\label{eqn:BTZnull}
ds^2 = \frac{\ell^2}{Z^2}\left[\frac{\ell^2}{UV}\left(1+  \Delta^2 \Sigma^2 (Z/\ell)^4\right) \, dU\, dV + Z^2 \left(\frac{\Sigma^2}{U^2} \, dU^2 + \frac{\Delta^2}{V^2} \, dV^2 \right) + dZ^2\right].
\end{equation}
The transformation from $(U_P,V_P,Z_P)$ to $(U,V,Z)$ is then
\begin{align}
\label{Carlip}
U_P &= -\ell  \frac{1-\Sigma\Delta (Z/\ell)^2}{1+\Sigma\Delta (Z/\ell)^2}\, \left(-U/\ell\right)^{2\Sigma}, \cr
V_P &= \ell \frac{1-\Sigma\Delta (Z/\ell)^2}{1+\Sigma\Delta (Z/\ell)^2 } \, \left(V/\ell\right)^{2\Delta}, \cr
Z_P &= \frac{2\sqrt{\Sigma\Delta} \, Z}{1+\Sigma\Delta (Z/\ell)^2}\,  \left(-U/\ell\right)^\Sigma \left(V/\ell\right)^\Delta .
\end{align}
Combining \eqref{Carlip} with \eqref{PtoFG} and using the identifications~$\Delta = 2\gamma_L$ and~$\Sigma = 2\gamma_R$ yields the transformation between the BTZ coordinates $U,V,Z$ and $(u,v,z)$.
In particular, the result is well-behaved in the extremal limit~$\Delta = 0$,~$\Sigma=1/2$, where it becomes
\begin{align}
\label{Untrans}
U(u,v,z) &= \ell \kappa u \left[1 - \frac{2(1-\kappa^2 uv)(\kappa z)^2}{4 \kappa^2 uv(1- \kappa^2 uv)-\kappa^2 uv(\kappa z)^2 }\right], \cr
V(u,v,z) &= \ell \kappa v \, \exp\left[-(\kappa z)^2\, \frac{8(1-\kappa^2 uv)^2-2(2- \kappa^2 uv)(1- \kappa^2 uv)(\kappa z)^2}{(4(1- \kappa^2 uv)-(\kappa z)^2)(4 \kappa^2 uv(1- \kappa^2 uv)-(2- \kappa^2 uv)(\kappa z)^2)}\right], \cr
Z(u,v,z) &= \ell \sqrt{\frac{-16(1-\kappa^2 uv)^3 (\kappa z)^2}{(4(1-\kappa^2 uv)-(\kappa z)^2 )(4 \kappa^2 uv(1- \kappa^2 uv)-(2-\kappa^2 uv)(\kappa z)^2)}}.
\end{align}
A non-degenerate (but rather implicit) transformation from Fefferman-Graham coordinates $u,v,z$ to the global coordinates $\tau, R, \theta$ of footnote \ref{global}  may then be obtained by combining eqns \eqref{Untrans} with the transformation
\begin{align}
\label{Uglobal}
U(\tau,R,\theta) &= -\ell \, \frac{2R\cos\theta+(1+R^2)\cos\tau}{2R(\cos\theta-\sin\theta)+(1+R^2)(\cos\tau+\sin\tau)}, \cr
V(\tau,R,\theta) &= \ell\exp\left[\frac{1}{2}+\frac{2R \sin\theta+(1+R^2)\sin\tau}{4R\cos\theta+2(1+R^2)\cos\tau} + \frac{2R\sin\theta-(1+R^2)\cos\tau}{2R(\cos\theta-\sin\theta)+(1+R^2)(\cos\tau+\sin\tau)}\right], \cr
Z(\tau,R,\theta) &=  \ell \frac{1-R^2}{\sqrt{[2R\cos\theta+(1+R^2)\cos\tau][2R(\cos\theta-\sin\theta)+(1+R^2)(\cos\tau+\sin\tau)]}}.
\end{align}
which relates the BTZ coordinates $U,V,Z$ to global coordinates $\tau,R,\theta$. We have used this procedure to generate figures~\ref{f:global} and~\ref{f:FG}.  In particular, we were able to identify the bulk horizons~$H^\pm$ in Fefferman-Graham coordinates from~\eqref{Untrans} and the known horizons in BTZ coordinates. We also identified singular surfaces of the transformation by examining the Jacobian; these are the plane~$v = 0$ and the dark surface labeled CS in figure \ref{f:FG}.  The CS surface was numerically mapped to the global coordinates by using equations~\eqref{Uglobal} and is shown in figure~\ref{f:global} as a set of lines, each of which is a contour of constant~$uv$. Note that the surface defined by these lines ends abruptly in the middle of the spacetime.  This edge corresponds to the boundary singularity at~$\kappa^2 uv = 1$.  Further examination reveals that the CS surface contains a set of branch points in the transformation to Fefferman-Graham coordinates. To make the coordinate transformation one-to-one, we must introduce an appropriate branch cut.  Our choice is indicated in figure \ref{f:FG}.  The transformation is one-to-one in the region between the cut, the boundary, and the CS surface.  In terms of our global coordinates (figure \ref{f:global}), the cut is a surface that starts near the internal edge of the CS surface, runs to the right above the CS surface, and terminates at $\mathscr{I}^\pm$.



\end{document}